\documentclass[applsci,review,accept,moreauthors,pdftex,10pt,a4paper]{mdpi} 

\usepackage{wasysym}

\usepackage{tabularx}
\usepackage{array}
\newcolumntype{L}[1]{>{\raggedright\let\newline\\\arraybackslash\hspace{0pt}}m{#1}}
\newcolumntype{C}[1]{>{\centering\let\newline\\\arraybackslash\hspace{0pt}}m{#1}}
\newcolumntype{R}[1]{>{\raggedleft\let\newline\\\arraybackslash\hspace{0pt}}m{#1}}
\newcolumntype{M}[1]{>{\centering\arraybackslash}m{#1}}
\newcolumntype{N}{@{}m{0pt}@{}}

\firstpage{1} 
\pubvolume{xx}
\issuenum{1}
\articlenumber{5}
\pubyear{2019}
\copyrightyear{2019}
\history{Received: date; Accepted: date; Published: date}
\updates{yes} 
\usepackage{upgreek}

\Title{Astrophotonic Spectrographs}

\Author{Pradip Gatkine $^{1,}$*\orcidA{}, Sylvain Veilleux $^{1}$ and Mario Dagenais $^{2}$}
\AuthorNames{Pradip Gatkine, Sylvain Veilleux, Mario Dagenais}

\address{%
$^{1}$ \quad Department of Astronomy, University of Maryland College Park, College Park, Maryland 20742, USA; veilleux@astro.umd.edu\\ 
$^{2}$ \quad Department of Electrical and Computer Engineering, University of Maryland College Park, College Park, Maryland 20742, USA; dage@ece.umd.edu} 

\corres{Correspondence: pgatkine@astro.umd.edu}

\abstract{Astrophotonics is the application of photonic technologies to channel, manipulate, and disperse light from one or more telescopes to achieve scientific objectives in astronomy in an efficient and cost-effective way. Utilizing photonic advantage for astronomical spectroscopy is a promising approach to miniaturizing the next generation of spectrometers for large telescopes. It can be primarily attained by leveraging the two-dimensional nature of photonic structures on a chip or a set of fibers, thus reducing the size of spectroscopic instrumentation to a few centimeters and the weight to a few hundred grams. A wide variety of astrophotonic spectrometers is currently being developed, including arrayed waveguide gratings (AWGs), photonic echelle gratings (PEGs), and Fourier-transform spectrometer (FTS). These astrophotonic devices are flexible, cheaper to mass produce, easier to control, and much less susceptible to vibrations and flexure than conventional astronomical spectrographs. The applications of these spectrographs range from astronomy to biomedical analysis. This paper provides a brief review of this new class of astronomical~spectrographs.}

\keyword{astrophotonics; arrayed waveguide gratings; echelle gratings; single-mode; near-IR; photonic lanterns}

\begin{document}

\section{Introduction}
The field of photonics has become an indispensable component of modern-day communication technology. The developments and demands from the telecommunication industry has driven a major boost in photonic technology and vice-versa in the last 35 years. Both fundamental and applied sciences have benefited from and contributed to the age of photonics. The platform of guided light in fibers and waveguides has opened the gates to next-generation instrumentation in astronomy. At the same time, the spin-offs from astrophotonics are extensively being adapted in other streams of science and technology. 

The field of astrophotonics revolves around collecting astronomical light into guided channels, manipulating the transport and reconfiguration of the light, and filtering/dispersing/combining the guided light. A combination of one or more of these functionalities has led to a wide spectrum of astrophotonic instruments. They serve the purpose of leading new astronomical investigations or making the current investigations more efficient and cost-effective. Some of these scientific frontiers are: Detailed kinematic studies of galaxies \cite{bland2016galaxy}, observing faint sources in the early universe \cite{stark2016galaxies}, discovering and characterizing exoplanets and their atmospheres \cite{des2002remote}, and studying the first galaxies in the universe \cite{bromm2011first}. Following these pursuits requires large telescopes and innovative ways to manipulate the light to extract the most information from a limited number of photons.

The next two decades will mark the age of thirty-meter scale telescopes \cite{stone2017development, ramsay2018instrumentation}. However, the volume, mass, and cost of conventional optical instruments grow as $D^{2}$--$D^{3}$ 
 (where $D$ is the diameter of the telescope), creating new challenges for instrumentation \cite{bland2006instruments}. At the same time, there is an explosion in the number of large astronomical surveys discovering a multitude of new transients, exoplanets, and galaxies. This necessitates development of new instruments for large telescopes that are compact and cost-effective while providing the flexibility to address the challenge of high-throughput characterization for these large surveys.   

\begin{figure}[H]
\centering
\includegraphics[width=14 cm]{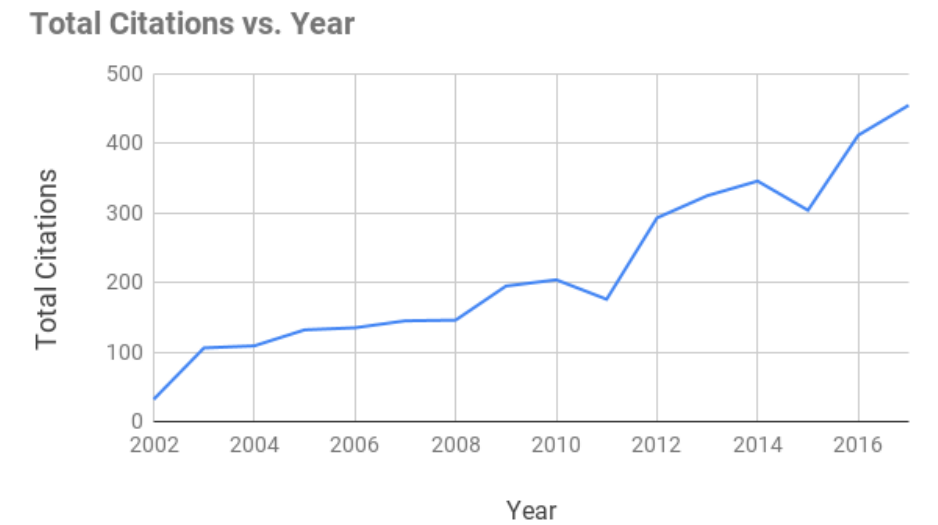}
\caption{Total number of new citations to papers in the field of astronomical spectroscopy using photonics. The publication data was obtained from the Web of Science interface (\url{http://apps.webofknowledge.com}). \label{Fig:Citations}}
\end{figure}

Photonic technologies provide a promising platform for building next-generation instruments that are flexible (in terms of light manipulation), compact (volumes of a few tens of cubic centimetres) and lightweight (a few hundreds of grams), thanks to manipulation of guided light \cite{allington2010astrophotonic, bland2009astrophotonics}. 
In addition, they are cost-effective, due to the advantages of mass-fabrication. Therefore, the astronomical community has pursued this direction very positively and built a wide variety of instruments from the near-ultraviolet (NUV) \cite{tuttle2010fireball} to mid-infrared (MIR) wavebands \cite{arriola2017mid}. As an example of the growth of astrophotonics, Figure $\ref{Fig:Citations}$ shows the number of citations of papers related to astrophotonic spectrographs. However, the current technical know-how is still far from complete or ideal, leaving many challenges that will likely be addressed in the near future. 

In this paper, we focus on the recent developments in the field of astrophotonic spectrographs. To paint a complete picture, in Sections \ref{sec2} and \ref{sec3}, respectively, we briefly discuss the steps to channel the light from the telescope to the spectrograph and condition it. Photonic dispersion concepts, and their implementation and challenges are described in Section \ref{sec4}. In Section \ref{sec5}, we discuss new developments in calibration and detection that are particularly of interest to astrophotonic spectrographs. Lastly, in Section \ref{sec6}, we summarize and discuss the impact of photonic spectrographs in the near future in astronomy as well as other applications. 

\section{Guiding the Light} \label{sec2}
The developments in the photonic industry and fabrication techniques in the last two decades have led to diverse adaptations of photonics optimized for the photon-starved field of astronomy. In~order to use  photonic advantage, the first step is to focus the light collected by the telescope onto a fiber, which is then used to efficiently carry the light to the spectrograph or other instruments.  

 \subsection{Light Pipes}
 Historically, one of the first applications of these photonic dividends 
  from communication industry came in the form of fiber fed (conventional) spectrographs \cite{lund1984prospects}. The simple concept of using an optical fiber to carry the light from the focal plane of the telescope to the spectrograph slit imparted immense flexibility to astronomical spectroscopy in several aspects: 
a) Size: compact solution for multi-object spectroscopy; 
b) modularity: the use of fibers isolates the spectrograph from the telescope,  enabling quicker change of instruments;
c) mechanical stability: due to isolation of the spectrograph, its alignment is more stable and immune from changes in the telescope orientation;
d) improved image stability due to image scrambling that occurs due to thousands of uncorrelated total internal reflections;
e) improved spectral calibration due to consistency of the illumination pattern between the source, flat field, and calibration sources.
 
These features of fiber feeding led to significant developments in multi-object spectroscopy and fiber-fed integral field units. The fiber feeding was limited almost exclusively to optical and near-IR wavebands 
(0.3--2.4 $\upmu$m) 
 due to the absorption properties of fused silica which is the principal material used in optical fibers \cite{labadie2016astronomical}. Recently, fibers with new materials such as lithium niobate and fluoride glasses (e.g., ZBLAN glass), and chalcogenide glasses (e.g., gallium lanthanum sulphide glass) are being developed in orde to extending this range to 
 10 $\upmu$m 
  for space-based applications \cite{labadie2016astronomical}. On the other hand, NUV guiding ($\sim$190--300 nm) has been enabled by high OH-content fibers \cite{oto2001optical}. 

A key aspect in the design of fibers to capture the focused light is the number of modes supported by the fiber, which in turn depends on the refractive index gradient and the cross section of the fiber. The astronomical light received at the focus of a ground-based telescope is seeing-limited for optical and near-IR wavebands (i.e., the FWHM of the spot size is governed by atmospheric turbulence and not the diffraction limit). Capturing the large spot size with maximal coupling efficiency requires the use of large multimode fibers (MMFs, $\sim$several tens of modes). Although they act as \textit{light pipes}, they are not suitable for any photonic manipulation of the light (such as filtering, interferometry, and spectroscopy in the photonic realm). The key reason is that most of these technologies, primarily inherited from the telecommunication industry, are developed for single-mode fibers (SMFs) which only propagate the fundamental mode, unlike MMFs which propagate multiple modes with different refractive indices.  


\subsection{Single-Mode Coupling}
Single-mode operation is critical for certain precision measurements, such as exoplanet radial velocity \cite{udry2007decade, halverson2016photonic, bechter2018radial}. With the advent of adaptive optics (AO) to correct for atmospheric aberrations, the spot size is reduced (i.e., higher Strehl ratio) to allow few-mode fibers to capture the focused light~\cite{horton2007coupling, corbett2009sampling}. Moreover, coupling the AO-corrected light on a SMF has recently been demonstrated using a lenslet array of 
 off-axis parabolic mirrors with a coupling efficiency of $\sim$25\% in the Y-band (970--1070 nm) \cite{bechter2016sky} and in the R-band (570--720~nm) \cite{garcia2016vision}.

With recent developments in extreme AO in concert with phase-induced apodized aperture (PIAA) lenses, a coupling efficiency as high as $\sim$74\% has been demonstrated in the H-band (1500--1600~nm) for coupling the focused light into a single-mode fiber \cite{guyon2003phase, jovanovic2016efficiently, corbett2005coupling}. This is close to the maximum achievable coupling efficiency in such a system with apodization in the diffraction limit ($\sim$90\%). Despite these developments, it is more convenient and economical to capture the light in a bundle of few-mode fibers and use the concept of photonic lanterns \cite{leon2005multimode, leon2013photonic} 
 to transition the few-mode fibers to SMFs \cite{horton2007coupling}, since at present extreme AO is effective for only bright astronomical sources in the NIR bands (e.g., i-band magnitude of <9). 

\subsection{Photonic Lanterns}
 The concept of photonic lanterns emerged from the need to efficiently combine the light-capturing efficiency of MMFs and the light manipulation flexibility of SMFs \cite{leon2005multimode}. A photonic lantern is essentially a MMF adiabatically tapering into a set of SMFs, allowing a low-loss transition. For highest efficiency, the number of SMFs should be the same as number of modes supported in the MMF, but fewer SMFs could be used if the efficiency requirement is lower \cite{leon2013photonic, birks2015photonic}. The technology has since matured to lead to several variants. The key categories include: a) A MMF to a bundle of SMFs, with a demonstrated efficiency of $\sim$85--90\% in the H-band \cite{noordegraaf2009efficient, birks2012photonic, trinh2013gnosis}; b) a MMF to a multi-core fiber with an array of identical SMFs \cite{birks2012photonic}; and c) an integrated photonic lantern, in which the transition from a multimode waveguide to many single-mode waveguides is inscribed in a glass block using ultra-fast laser inscription (ULI) \cite{thomson2011ultrafast, cvetojevic2017modal}. While initially produced in the J (1100--1300 nm) and H bands (1400--1700 nm), photonic lanterns are now being developed in the mid-infrared bands as well using femto-laser inscription \cite{arriola2014new}.  
 
In addition to conveniently allowing single-mode functionality with the astronomical light, photonic lanterns also make it possible to reformat the beam in many ways by rearranging the output single-mode fibers (e.g., in a slit form or closed packing \cite{birks2015photonic} ) for the next instruments in the pipeline (e.g., a diffraction-limited spectrograph or an interferometer) \cite{spaleniak2013integrated, gatkine2018towards, harris2018nair}.  


\section{Filtering and Combining the Light}\label{sec3}
Before the captured and guided light (in SMFs) is fed to a photonic spectrograph, it can be further conditioned depending on the scientific objective of the astronomical observation as well as the waveband of interest. In this section we discuss two major types of conditioning relevant to astrophotonic spectroscopy.

\subsection{OH-Emission Suppression}
Ground-based NIR spectroscopy faces the problem of bright NIR background due to a series of narrow ($\sim$0.1 nm) emission lines (telluric lines) from atmospheric hydroxyl (OH) molecules~\cite{meinel1950oh}. Background subtraction is very difficult, due to short-timescale fluctuations in the brightness of these lines, making observations of faint sources impractical. This problem is pronounced in the J (1100--1300 nm) and H bands (1400--1700 nm). Sending this background-dominated light to a low/moderate resolution spectrograph (with $\delta \lambda$ > 0.05 nm) leads to scattering of the bright background which is difficult to remove after the fact. An improved background subtraction requires suppression of these lines before it interacts with the dispersing optics. A promising way to suppress these emission lines before dispersion with high rejection ratio and low loss is using Bragg gratings in single-mode fibers \cite{trinh2013gnosis}. 

Bragg gratings act via variations in the refractive index, leading to constructive interference of the backward propagating wave of a certain frequency, thus that frequency is reflected backwards. A complex grating reflects a desired profile in wavelength, i.e., a filter can be constructed using this concept. Such a filter can be implemented in four ways: a) Individual SMFs \cite{bland2011complex}, b) a multi-core bundle of single mode fibers \cite{lindley2014demonstration, bland2016multicore}, c)
 on-chip waveguide Bragg gratings \cite{zhu2016arbitrary}, and d) 3D photonic Bragg gratings in a glass block using ULI \cite{brown2012ultrafast}. So far, Bragg gratings on individual SMFs has been demonstrated on sky with $\sim$100 notches, with a rejection ratio of $\sim$1000 at a notch-width $\delta \lambda$~$\sim$~0.15~nm~\cite{trinh2013gnosis, ellis2018praxis}. 

\begin{figure}[H]
\centering
\includegraphics[width=15.5 cm]{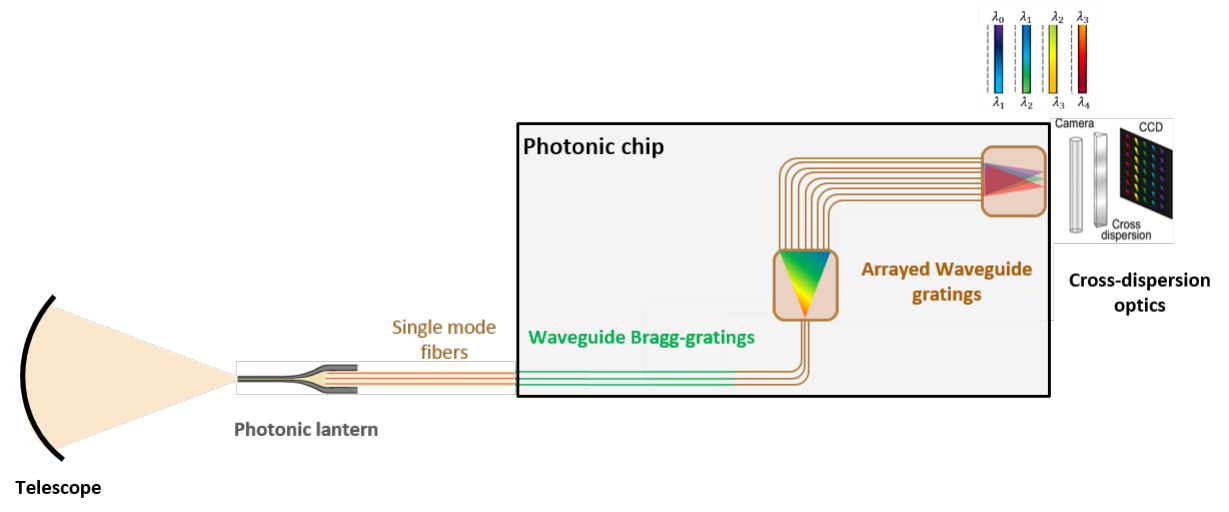}
\caption{A schematic of the full setup of an integrated photonic spectrograph with OH-emission suppression, reprinted with permission from \cite{gatkine2018towards}. \label{Fig:AWG_full_path}}
\end{figure}

A key challenge of creating notches at precise locations for on-chip and fiber-based Bragg-gratings is the need to control temperature, since the refractive index is a sensitive function of temperature~\cite{lindley2016post}. It is easier to control the temperature on a compact chip than on extended fibers. With on-chip Bragg-gratings, however, there is an issue of polarization dependence with on-chip rectangular waveguides. Even with square waveguides, small dissimilarities in the upper and lower cladding can lead to shifts in the notch wavelengths. Hence, fabrication process control is critical. These shortcomings are currently being alleviated significantly by improving the fabrication recipes and device modeling \cite{zhu2016arbitrary, zhu2016ultrabroadband, hu2018characterization}. Considering the potential of combining the filtering and dispersion steps on the same chip (as shown in Figure \ref{Fig:AWG_full_path}), this line of ideas is worth exploring.  

On-chip ring resonators offer another way to suppress the telluric lines, where each line is suppressed by an individual ring \cite{ellis2017photonic}. However, due to small ring radii required (e.g., $<$ 10 $\upmu$m in the NIR band), the bending losses tend to be high. This necessitates the use of high index contrast platforms (e.g., $Si$ on $SiO_{2}$), which, due to high confinement of the modes, face low coupling efficiency with SMFs. This problem can be alleviated with the use of mode-matching tapers. 

\subsection{Interferometry}

In certain astronomical studies, such as measuring the stellar diameters and observing exoplanets, it is crucial to achieve high-resolution imaging and/or suppress light from a bright star next to a faint planet. For ground-based observations, the necessity of high spatial resolution requires combining light from two or more telescopes. In the past, the technique of interferometry has been used to measure stellar radii using two or more telescopes separated by a distance D, to allow $\sim$$\lambda/D$ angular resolution \cite{coude1992fluor}. Another way, called aperture masking interferometry, achieves the diffraction limit through artificially created sub-apertures formed by segmenting the telescope pupil \cite{tuthill2000michelson, jovanovic2012starlight}. The image (i.e., brightness distribution) is reconstructed using Van Cittert--Zernike theorem, which states that the spatial coherence function of light is the Fourier transform of the angular brightness distribution of the light source \cite{roychoudhuri1995van}. Integrated photonic reformatters and interferometers have recently been developed with compact designs, better phase control, and mechanical/thermal stability \cite{tuthill2010photonic}.

\begin{figure}[H]
\centering
\includegraphics[width=15.5 cm]{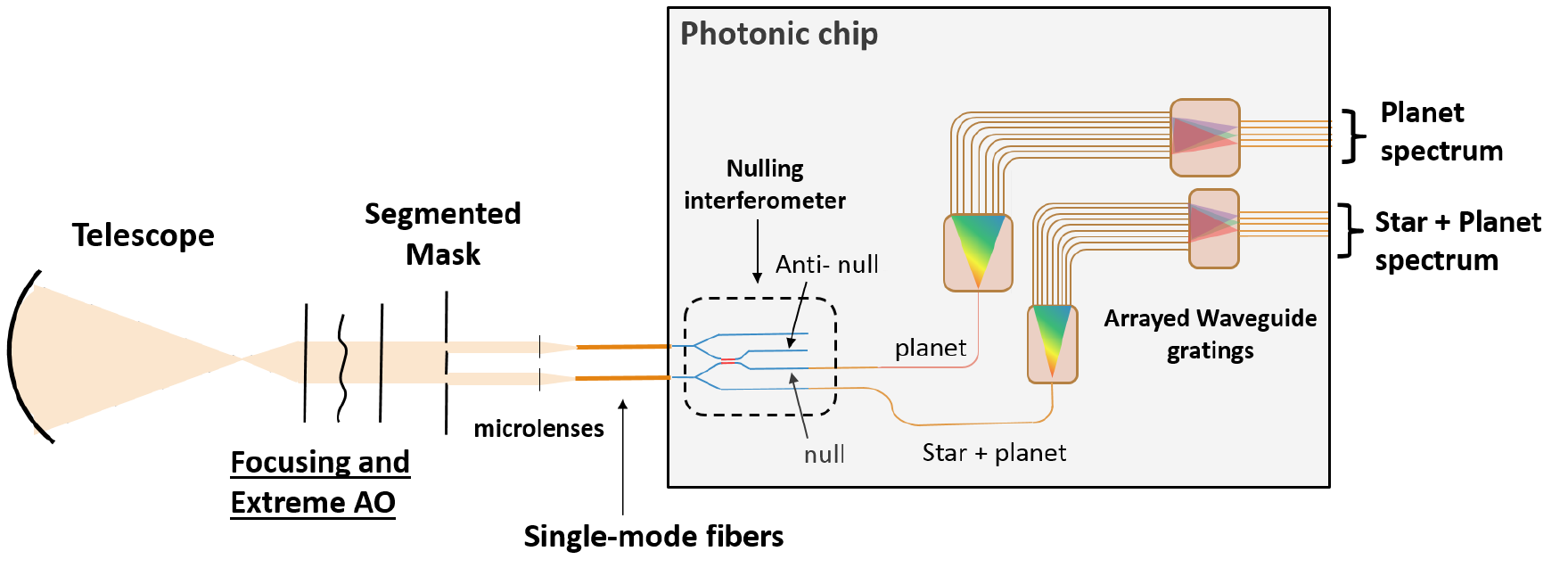}
\caption{A simple schematic of the full setup of an integrated photonic spectrograph specifically designed for low-resolution exoplanet spectroscopy by integrating nulling interferometer and AWGs on one chip. The two arms of the nulling interferometer have a phase difference of $\pi$. The two AWGs are fed the outputs of the nulling interferometer to acquire separate spectra of star$+$planet and planet only (coming from the null output). Additional processing of the spectra will be required to eliminate residual starlight (say, from diffraction speckles) in the planet only spectrum.   \label{Fig:AWG_full_path_null}}
\end{figure}

A nulling interferometer is an offshoot of the technique described above, where an extra phase delay of $\pi$ is added to one of the two combining beams to create destructive interference of the central source (e.g., a bright star), making any off-center source visible (e.g., a planet). A photonic nulling interferometer called GLINT has recently been demonstrated which can be used for observing  exoplanets, possibly able to measure their spectra in the future \cite{lagadec2018glint}. A current analogue which offers these functionalities (R~$\sim$~20, 500, 4000 and angular resolution of tens of milliarcseconds in the K-band) is the GRAVITY instrument---a VLT interferometer \cite{gillessen2010gravity, abuter2017first, abuter2018detection}. A photonic counterpart of GRAVITY will be shoebox-sized, with improved mechanical stability, thus allowing deeper observations of faint, background-limited sources. A sample schematic of such a future device is shown in Figure \ref{Fig:AWG_full_path_null}. 

Integrated photonic interferometers also show promise in the mid-IR wavelengths, with ongoing developments in mid-IR fabrication technologies \cite{diener2017towards}. 

All of these form the set of next-generation photonic devices that will illuminate the next-generation photonic spectrographs and perhaps be integrated with them on a monolithic~platform. 

\section{Dispersing the Light}\label{sec4}

The astronomical light guided in the single mode fibers can be dispersed in many ways, most of which are inherited from the telecommunication technique wavelength division multiplexing (WDM). This technique emerged from the need to increase the data density of the existing fiber-optic networks by using multiple wavelengths to carry multiple data packets simultaneously. Thus, a variety of photonic platforms were developed to combine different wavelengths (multiplex) and disperse them (de-multiplex). We will discuss the on-chip photonic dispersion platforms that are most relevant for astronomical spectroscopy and will potentially be commissioned on telescopes in near future. 

\subsection{Arrayed Waveguide Gratings (AWGs)}

\noindent
\subsubsection{The Concept}
\begin{figure}[H]
\centering
\includegraphics[width=14 cm]{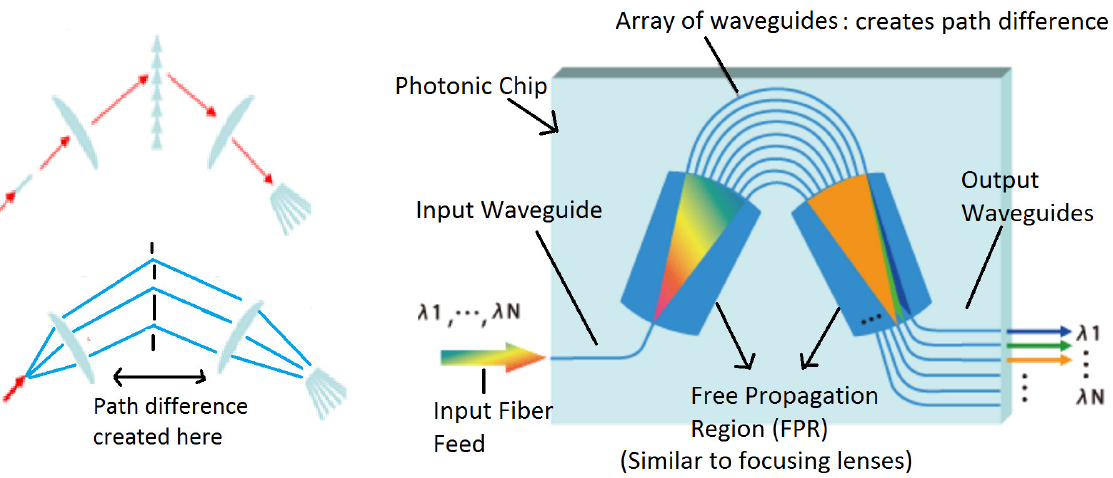}
\caption{Analogy between a conventional grating spectrograph (left) and arrayed waveguide gratings (right), reprinted with permission from \cite{gatkine2016development}. 
\label{Fig:AWG_analogue}}
\end{figure}

Arrayed waveguide gratings 
 are a widely used technology in the telecommunication industry and a direct photonic analogue of a conventional diffraction grating spectrograph, as shown in Figure~\ref{Fig:AWG_analogue}.
 The concept of AWGs, first presented in 1988, is based on a phased array of channels/antennae in radio receivers and transmitters \cite{smit1988new, leijtens2006arrayed}. A single-mode fiber couples to a single-mode waveguide on the chip. The input waveguide illuminates an array of single-mode waveguides through a slab waveguide, called a free propagation region (FPR). Each waveguide in the array has an identical path difference, $m \lambda_{0}$ with respect to its adjacent waveguide, where $m$ is the central spectral order and $\lambda_{0}$ is the central wavelength. The emergent light interferes in the output FPR and forms constructive interference fringes of different wavelengths within a spectral order (where free spectral range $\sim$$\lambda/m$) at different locations on the Rowland circle. Thus, the wavelengths within a spectral order are dispersed horizontally. The discrete resolving elements can then be sampled by output waveguides. 

However, different spectral orders overlap at the focal plane (see Figure \ref{Fig:AWG_full_path}). Essentially, all wavelengths the same value of $m\lambda$ will overlap. Hence, for a complete spectrum, the spectral orders need to be separated using a cross disperser (i.e., an order sorter) or a set of secondary AWGs (called tandem AWGs) \cite{takada200110}. For continuous sampling of wavelengths, as is often required in astronomical spectrographs, the chip is cleaved at the focal plane and cross-dispersed in the perpendicular direction to get a full spectrum with the orders separated vertically (see Figure  \ref{Fig:AWG_full_path}). In the design of AWGs, especially at high spectral resolution ($R$), the AWG does the heavy-lifting with finer dispersion ($\delta\lambda$~$\sim$~$\lambda/R$), and the cross-dispersive optics do a coarse dispersion ($\Delta \lambda$ $\sim$ $\lambda/m$) \cite{cvetojevic2012first}.

\subsubsection{Fabrication}

Photonic AWG is the most explored on-chip dispersion technique. Availability of electron-beam lithography has allowed imprinting nanoscale structures on silicon-based platforms. In the last few years, silicon nitride on silica ($Si_{3}N_{4} / SiO_{2}$) has emerged as a low-loss and high index contrast material for photonic devices \cite{bauters2011ultra, dai2011low}. Single-mode propagation loss of less than 1 dB/m has been demonstrated at 1550 nm with wide waveguide geometries \cite{huffman2018integrated}. At the same time, high index contrast allows for strongly guided modes, thus smaller bend radii ($r_{bend}$ $\sim$ 2 mm) are possible with low bending losses and low cross-talk among the waveguides in the array.

However, a key drawback with high index contrast waveguides (i.e., strongly guided modes) is that they are difficult to couple with SMFs, due to a large mismatch in the mode shape and size. Recent work at the University of Maryland has demonstrated $\sim$95\% coupling efficiency with high numerical aperture SMFs at 1550 nm by using an adiabatic taper-down profile and a rectangular region of appropriate width added to the waveguide before it is butt-coupled with the fiber \cite{zhu2016ultrabroadband}. Application of this technique has produced a moderate-resolution (R $\sim$ 1500), wide-band (1450--1650 nm) AWG with $\sim$25--30\% end-to-end throughput (fiber-chip-fiber) at 1600 nm \cite{gatkine2016development, gatkine2017arrayed}. With more improvements underway, the $Si_{3}N_{4} / SiO_{2}$ platform has a lot to offer towards building high-throughput astrophotonic instruments on a chip. 

Another technique called ultra-fast laser inscription (ULI) or femto-second laser direct writing and is used for writing micron-scale structures in glass substrates \cite{choudhury2014ultrafast}. ULI-based AWG has been demonstrated for the first time this year \cite{douglass2018femtosecond}. Thanks to the one-step fabrication process, this is a quick, scalable method with easier process control or repeatability. This technique has been demonstrated for building optical, NIR \cite{macdonald2010ultrafast}, and mid-IR devices (using chalcogenide glass substrates) \cite{butcher2018ultrafast, thomson2009ultrafast}. ULI allows fabrication of a 3D structures in glass without introducing additional steps, which opens a great avenue for integrating photonic lanterns, pupil remappers, beam combiners, and/or AWGs in a single glass block.  However, the low-index contrast ($\sim$$10^{-3}$) of ULI means the waveguide separation needs to be large to prevent crosstalk and the bending radii need to be large ($\sim$25 mm) to avoid high bending losses. This increases the size of the device, which could cause unwanted systemic variations. Despite these issues, with rapid advancements in this field, this versatile technique shows great promise for integrated photonic devices over wide wavebands. 

\subsubsection{Challenges and the Future} 
While the concept of AWGs has matured over the last three decades, there are some specific issues that need to be addressed in the context of astrophotonic AWGs. The most important requirements are high-throughput over broad bands, polarization independence, and high coupling efficiency with SMFs. Solutions to these issues exist in standalone devices or material platforms \cite{stoll2017high, zhu2016ultrabroadband}. It is critical to bring these together to build an astrophotonic AWG spectrograph with better performance than existing conventional astronomical spectrographs, such as volume phase holographic (VPH) gratings.   

Astrophotonic AWGs have great potential to be deployed on space telescopes. They can work as low-resolution filters or high-resolution spectrographs. By stacking the AWGs and feeding light from different sections of an extended source to separate AWGs, a compact integral field unit (IFU) spectrograph can be constructed. Such a stack can also be combined with automated fiber positioners (e.g., \textit{starbugs} \cite{gilbert2012starbugs}) to make rapidly reconfigurable multi-object spectrographs.

\subsection{Echelle Gratings}

\subsubsection{The Concept}
\begin{figure}[H]
\centering
\includegraphics[width=14 cm]{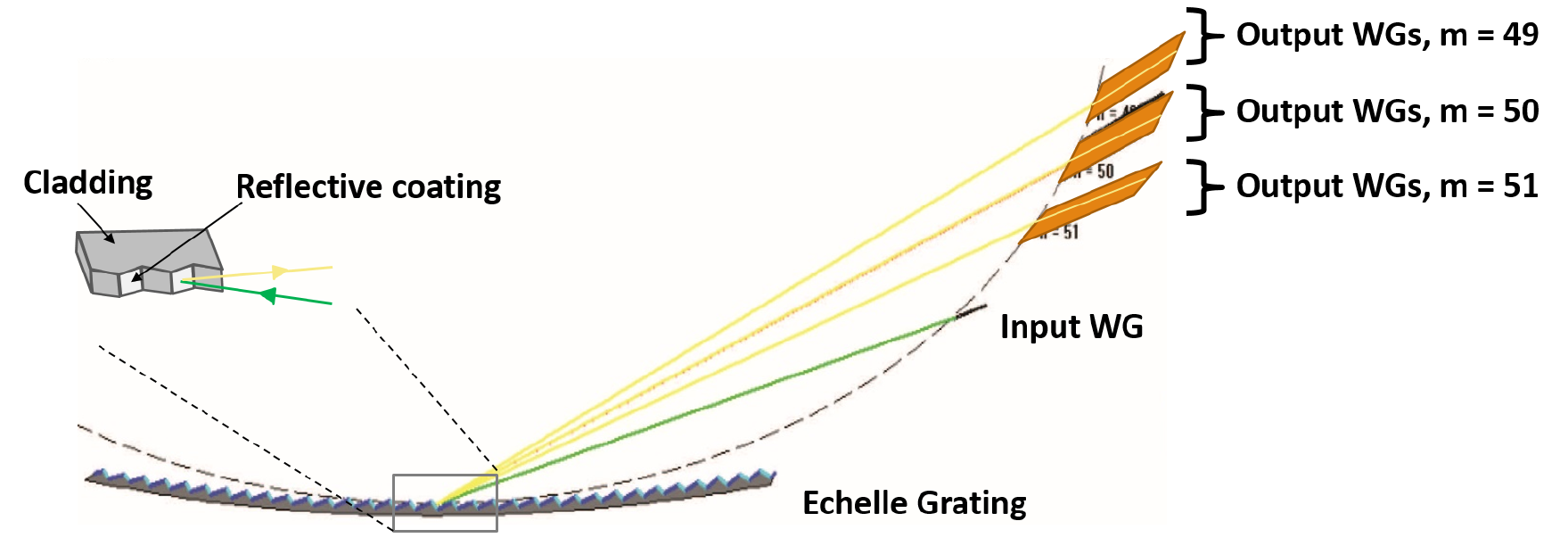}
\caption{Schematic of a photonic echelle grating spectrograph. A ray representation is shown for simplicity although the dimensions of the waveguides and reflecting surfaces require a wave treatment. Green shows the light emanating from the input waveguide and yellow  shows the reflected light. The brown colored region shows a number of output waveguides sampling the dispersed light from different spectral orders (for e.g., orders 49, 50, and 51). \label{Fig:Echelle_analogue}}
\end{figure}

Another photonic technique 
 based on a conventional spectrograph is photonic echelle grating (also called planar concave grating) spectrograph. A sample schematic is shown in Figure \ref{Fig:Echelle_analogue}. The light from a single mode fiber is coupled with the input waveguide which then illuminates the grating through the free propagation region. The grating teeth are designed to have reflectivity, which creates constructive interference fringes for different wavelengths in a spectral order at different locations in the focal plane. The gratings can be designed to focus most of their light in a particular order by controlling the blazing angle. The geometry of echelle gratings can be designed in many ways, the most common being the Rowland circle method \cite{bland2006instruments}, the two-point stigmation method \cite{horst2009silicon}, and perfectly chirped gratings \cite{lycett2013perfect}. The high-efficiency reflection can also be achieved in a few different ways, such as by coating the vertical surface of grating teeth with reflective material like chromium or silver \cite{xie2018silicon, sciancalepore2015low, feng2011fabrication} or using distributed Bragg reflectors (DBRs) \cite{brouckaert2008planar}.

Photonic echelle gratings (PEGs) are particularly well suited for high-finesse (i.e., cross-talk~$\apprle$$-$25~dB) spectrographs, thanks to the much larger number of grating teeth that can be incorporated compared to the number of arrayed waveguides in an AWG. In addition, due to their compactness, multiple PEGs can be etched on a single compact chip in a tandem echelle design~\cite{bland2006instruments} without losing material uniformity. A small footprint also implies smaller propagation loss.

\subsubsection{Fabrication} Although the potential advantages of PEGs are attractive at face value, their fabrication of PEGs is a complex, multi-step process due to the requirement of reflective surfaces. In addition, the scale of the gratings requires the use of precision lithography, such as electron-beam lithography. A promising new technique uses electron-beam lithography to achieve a low-loss (insertion loss $\sim$1.4 dB) PEG using silver coating on the reflecting facet with a $Si_{3}N_{4}/SiO_{2}$ platform \cite{xie2018silicon}. Recently, low-loss PEGs have also been demonstrated in the J-band using a SiNOI platform \cite{sciancalepore2018band}. Another method for on-chip reflection, called DBR, is simpler for fabrication since there is no extra step to deposit reflective material. It has shown promise for astrophotonic echelle gratings \cite{brouckaert2008planar}, albeit with a higher insertion loss ($\sim$1.5 dB).

\vspace{1ex}
\noindent
\textbf{Challenges and future:} Echelle gratings with reflective facets have potential to be broadband, since the wavelength dependence of the design is minimal. A key challenge with echelle gratings is polarization independence. Both metal facet and DBR have directionality due to the fabrication process and inherent design, respectively. New designs are being proposed to tackle this problem (e.g., by introducing a polarization compensation area in the free propagation region \cite{zhu2008design}) and will be further developed in the near future.

\subsection{Fourier Transform Spectrograph}

\subsubsection{Concept} 
\begin{figure}[H]
\centering
\includegraphics[width=14 cm]{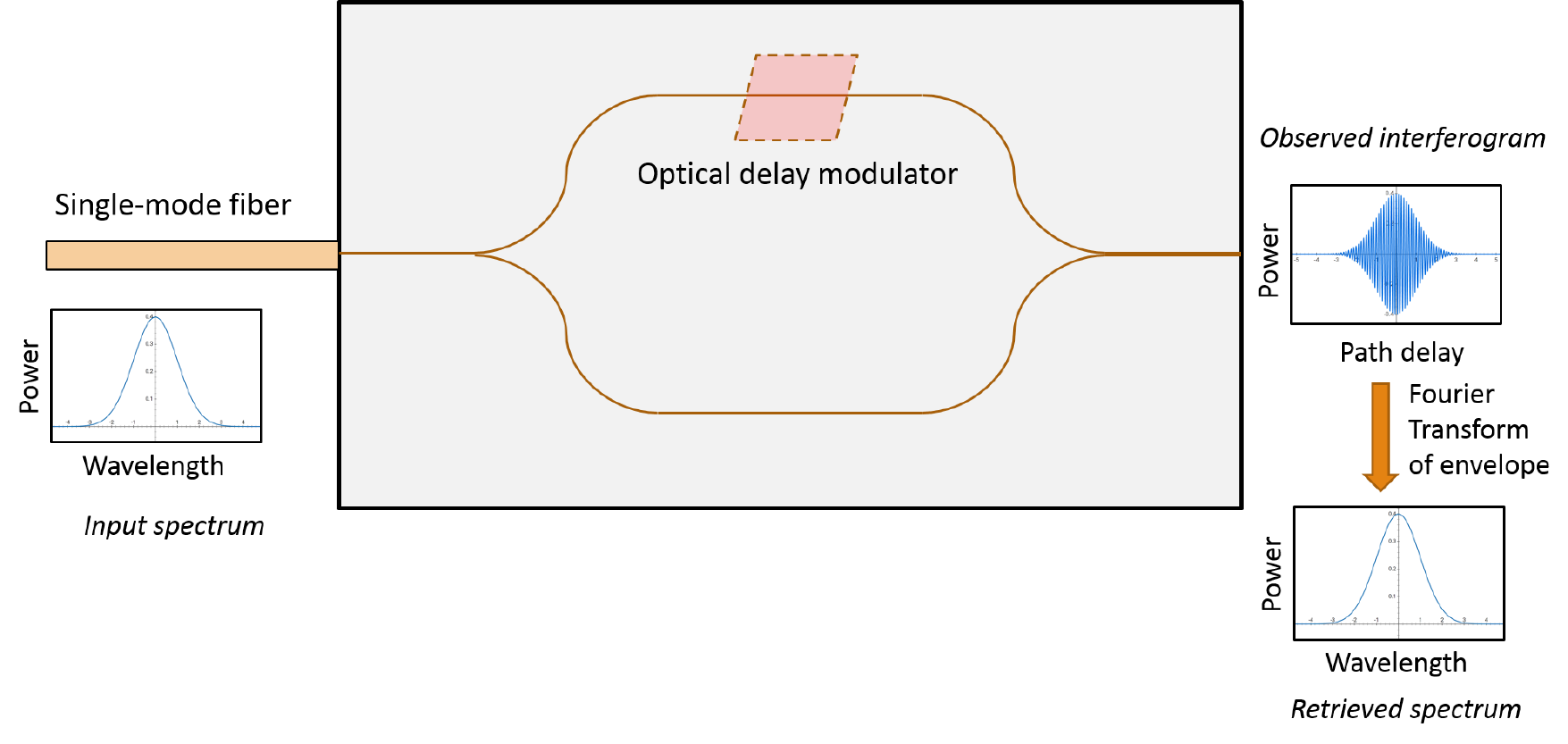}
\caption{Schematic of a photonic Fourier transform spectrograph. The input light from a single-mode fiber is coupled to the waveguide. This waveguide splits into two identical arms, with one fixed and the other with a variable path delay introduced (by modulating the refractive index). The two waveguides are combined and the interference pattern (called the interferogram) is measured at the output. The envelope of the interferogram is the Fourier transform of the input signal and is thus extracted in post-processing. A longer path delay is required to get a higher order interference fringe and thus a~higher spectral resolution. \label{Fig:FTS_schematic}}
\end{figure}

The concept of a Fourier transform spectrograph (FTS) has been around in astronomy for over half a century now \cite{connes1970astronomical, ridgway1984astronomical}. A conventional FTS is an adaptation of a Mach--Zehnder interferometer with a way to precisely control the optical path delay between two identical beams created using a beam splitter. Precise control of the path length necessitates an ultra-stable structure for the FTS. A~photonic FTS can easily resolve this issue by having a monolithic and miniaturized on-chip construction which, to a great extent, eliminates mechanical and thermal fluctuations. The schematic of a FTS is shown in Figure \ref{Fig:FTS_schematic}. The input fiber carrying the astronomical light is coupled with a waveguide which symmetrically splits into two identical waveguides (i.e., a 1 $\times$ 2 beam splitter). An optical delay modulator introduces a precise and known optical delay in one of the waveguides. The phase modulation is typically achieved using thermal \cite{harris2014efficient, souza2018fourier} or electro-optical modulation \cite{xu2005micrometre, subbaraman2015recent, dong2015nano}. The beams are combined and the power is measured using a single-element detector. By varying the path delay, an interferogram is constructed. The envelope of this interferogram is the Fourier transform of the incoming spectrum. Thus, the input spectrum is retrieved by inverse Fourier transform of the~envelope. 

Photonic FTSs, like their conventional analogues, have several advantages \cite{ridgway1984astronomical}. The FTSs are unique in terms of providing arbitrarily large spectral resolution ($\sim$10--100 m/s) over a broad band of wavelengths with excellent sensitivity. The downside is the integration time required for obtaining the spectrum. Unlike dispersive spectrometers, FTSs do not experience a degradation in the signal-to-noise ratio at a high spectral resolution, since the power is not distributed across small spectral channels \cite{harwit2012hadamard}. A convenient feature of FTSs is that they can offer flexible resolving power (governed by the extent of optical depths covered) depending on the scientific requirement, making them versatile. Moreover, for photonic FTSs, the detector can be a single-element detector directly butt-coupled to the photonic chip or channeled via a SMF. There is renewed interest in FTSs for exoplanet characterization, thanks to the plentiful discoveries of  exoplanets from the $Kepler$ mission \cite{borucki2010kepler, hodges2018fourier}. FTSs are also vital for measuring the radial velocities in binary star systems \cite{behr2009stellar}. Photonic single-mode FTSs are poised to provide a diffraction-limited, stable platform for performing sensitive high spectral resolution exoplanet studies. By using distinguishing spectral features (such as certain molecular lines in the mid-IR), an exoplanet spectral feature can be extracted from the interferogram without physically separating the star and planet light (such as in a nulling interferometer) \cite{schwartz2012spectroscopic}.   

\subsubsection{Fabrication}
Although the fabrication of FTSs is, in principle, very similar to that of AWGs, there are important differences due to addition of electronically controlled phase modulators. Currently,  silicon-on-insulator (SOI) is the prevalent platform for fabricating such devices where silicon is the guiding material and fabricated using standard complimentary metal-oxide-semiconductor (CMOS) processes. Various materials such as titanium nitride \cite{fang2011ultralow} or nickel--silicon \cite{van2010integrated} are used as heating elements and the silicon substrate and/or waveguide may be doped to introduce electrical conductivity \cite{harris2014efficient}. In electro-optical modulators, a p-i-n junction is embedded around the waveguide using standard CMOS processes to tune the refractive index.    

\subsubsection{Challenges and the Future }
Astrophotonic FTSs are in their infancy and will see a boost as other photonic technologies mature. There are several challenges that need to be resolved to enable astronomical spectroscopy with photonic FTSs, the most important being the throughput. The traditionally used SOI waveguides have higher propagation losses ($\sim$1--2 db/cm) due to the sharp index contrast between the Si core and SiO$_{2}$ cladding. In addition, doping and introduction of metals further degrade the transmission efficiency of the waveguides (by >0.5 dB). Recently, a photonic FTS has been demonstrated in a  more efficient silicon nitride platform (propagation loss $\sim$0.1 dB/cm) \cite{nie2017cmos}. Such developments show promise for boosting the throughput. Another important issue with FTSs is the birefringence and high dispersion of materials such as Si, SiO$_{2}$, and Si$_{3}$N$_{4}$ (in optical and near-IR). Since optical path modulation is achieved by modulating the refractive index, the optical path changes by a different amount for different polarizations as well as different wavebands. This creates a trade-off between the spectral resolution and bandwidth of the spectrograph. However, precise knowledge/calibration of the wavelength dependence of the thermal modulation can, in principle, alleviate this issue. Several new techniques have recently been demonstrated to effectively correct for temperature drifts and non-linearity of the thermal/electro-optic modulators~\cite{herrero2017temperature, souza2018fourier, kita2018high}. FTSs are also being demonstrated in the mid-IR range~\cite{nedeljkovic2016mid} ($\lambda$ $\sim$ 3.7 $\upmu$m, spectral resolution $\sim$5000), which is particularly useful for exoplanet characterization, as described above. All of these developments indicate a rapidly growing potential of photonic FTSs for astronomical spectroscopy.     


\section{Calibration and Detection}\label{sec5}
Precise wavelength calibration is crucial in spectroscopy, especially with high-resolution spectrographs (e.g., for radial velocity measurements with $\Delta v$ $\sim$ 10 cm/s), to ensure any short-term or long-term variations in the system are corrected. Currently, iodine cells are used for precision wavelength calibration, but it is difficult to get uniform wavelength coverage from them due to non-uniform absorption line density and saturated lines. Fabry--Perot interferometers and photonic combs solve this problem by creating high density of lines for monitoring the wavelength calibration precisely. Astronomical community has also started using frequency combs for calibration~\cite{wilken2012spectrograph}. Recently, photonic combs were used to remove the systematic aberrations of a wide-field spectrograph~\cite{bland2017mapping}. These new developments integrate well with the photonic spectrographs described above and are geared towards making the photonic spectrographs more stable and precise.

As described in Section \ref{sec4}, AWGs and PEGs are arguably the most promising and mature photonic dispersion technologies. For both, the focal plane is along a Rowland circle. Exposing the focal plane of the chip 
to free space for cross dispersion and/or imaging on the detector leads to significant Fresnel losses due to refractive index mismatch. On the other hand, a higher refractive index is desirable on-chip for better mode confinement, and lower propagation and bending losses. Avoiding the index jump to free space can improve the overall throughput of the spectrograph by as much as a factor of two. It should be noted that single-order or tandem AWGs/PEGs can be designed so that cross dispersion is not required. 

Therefore, from a throughput perspective, it is advantageous to directly couple the focal plane of these chips to the detector. This can either be a linear detector array or, in case of stacked photonic circuits, a two-dimensional detector array. Due to the curved focal plane, a curved detector is best suited for these devices \cite{harris2012applications}. A smaller pixel pitch ($\apprle$10 $\upmu$m) is also desirable to achieve the best spectral resolution in the direct coupling case, since most of the dispersed light is focused in an area of the order of a few hundred microns. Development of such detectors is underway and a successful concave CMOS detectors with a radius of curvature of 150 mm and pixel pitch of $\sim$7 $\upmu$m were reported only few months ago \cite{lombardo2018curved}. A highly curved CMOS detector (curvature radius $\sim$20--30 mm) by Microsoft has also been reported very recently with a pixel pitch <10 $\upmu$m \cite{guenter2017highly}. There have also been AWG designs with a flat focal plane \cite{lu2005design}, however, flatness of the focal plane is wavelength dependent. Therefore, for versatile uses, curved detectors will likely be a better solution.  


\section{Summary}\label{sec6}
Borrowing heavily from the telecommunication industry, the field of astrophotonic spectrographs has seen tremendous growth since its inception about two decades ago. Every year, new devices and improvements are being reported in every aspect of spectroscopic instrumentation discussed in this paper, i.e., guiding, filtering, combining, dispersing, and the detection of the light. In this paper, we have focused on three photonic spectroscopic techniques which show the best potential for astronomical spectroscopy---arrayed waveguide gratings, photonic echelle gratings, Fourier transform spectrographs. For each technique, we briefly discussed their relevance to astronomy, their underlying concept, fabrication methods, challenges, and their upcoming solutions in the context of astronomical spectroscopy. We have summarized various key aspects of these techniques in Table \ref{tab:summary}. To match the requirements of the photon-starved field of astronomy, efforts are being taken to achieve a throughput and performance comparable or even better than the existing conventional spectrographs, while miniaturizing the devices by orders of magnitude. In many aspects, the photonic approach is opening doors for new ways of processing the light, such as OH-emission suppression and active control/manipulation of optical path. These developments are well timed for the emerging era of thirty-meter scale terrestrial telescopes (e.g., TMT, ELT, and GMT). The next generation of space-based telescopes are also going to benefit greatly from the photonic dividends, thanks to the stability, compactness, and cost-effectiveness of photonic instrumentation.

For building astrophotonic spectrographs, it is important to address the challenges at hand, mainly from a fabrication perspective. In a broad context, the key improvements-in-waiting are: a)~Exploring new material platforms and stable processing technologies (over the full device size) for achieving high throughput over a broad band, b) developing polarization insensitive devices to cater to the unpolarized astronomical sources, c) integrating most of the devices in a spectrograph pipeline on a single chip to minimize the number of interfaces and subsequent loss of photons, and d)  detectors optimized for direct coupling with photonic chips. The developments described in this paper are not just limited to astronomy, but also applicable to a diverse set of fields, such as remote sensing, biophotonics, biomedical analysis, microfluidics, and the telecommunication industry. It is clear that the field of astrophotonics is gearing up towards building an ecosystem of instruments and spectrographs providing a diffraction-limited performance for the next generation of telescopes.  

\begin{table}[H]
\caption{A summary of astrophotonic spectrograph techniques showing demonstrated capabilities to achieve certain specifications/characteristics. It should be noted that the specifications mentioned here are exemplar to show the current state of the art. Particular device specifications vary according to science goals, wavelength range, material platform, fabrication methods, etc.} 
\label{tab:summary}
\begin{center}       
\begin{tabular}{C{3cm}C{3.5cm}C{3.5cm}C{3.5cm}} 
\toprule
\rule[-1ex]{0pt}{3.5ex} \textbf{Characteristic} & \textbf{AWG}  & \textbf{PEG}  & \textbf{FTS} \\
\midrule
\rule[-1ex]{0pt}{3.5ex}  Concept & Phased array & Blazed grating & Mach--Zehnder interferometer \\
\midrule
\rule[-1ex]{0pt}{3.5ex} Wavelength range demonstrated &  0.4--5.5 $\upmu$m \cite{ali2017compact, gatkine2017arrayed, malik2013germanium} & 0.8--4 $\upmu$m \cite{ma2013cmos, xie2018silicon, muneeb2013silicon} & 0.5--4 $\upmu$m \cite{podmore2018chip, velasco2013high, nedeljkovic2016mid} \\
\midrule
\rule[-1ex]{0pt}{3.5ex} High spectral resolution  & $\sim$1 $\times$ 10$^{4}$ \cite{cheben2007high} &  $\sim$1 $\times$ 10$^{4}$ \cite{ma2011spectrometer} & $\sim$1 $\times$ 10$^{5}$ \cite{velasco2013high}\\
\midrule
\rule[-1ex]{0pt}{3.5ex} Low crosstalk & $-30$ dB \cite{dai2011low} & $-30$ dB \cite{xie2018silicon} & N/A \\
\midrule
\rule[-1ex]{0pt}{3.5ex}  Low insertion loss & 1 dB \cite{gatkine2016development} & 1.4 dB \cite{sciancalepore2018band} & 1.2 dB \cite{kita2017chip}\\
\midrule
\rule[-1ex]{0pt}{3.5ex} Typical footprint  & $\sim$100 mm$^{2}$ \cite{gatkine2017arrayed} & $\sim$10 mm$^{2}$ \cite{xie2018silicon}  & $\sim$1 mm$^{2}$ \cite{souza2018fourier}\\
\midrule
\rule[-1ex]{0pt}{3.5ex}  Key advantages  & Easy fabrication, high throughput & Broadband, lower crosstalk  & Small footprint, high resolution \\
\midrule
\rule[-1ex]{0pt}{3.5ex}  Key challenges  & Cross dispersion, polarization dependence & Throughput, cross dispersion, polarization dependence & Wavelength-dependent path length, propagation loss\\

\bottomrule
\end{tabular}
\end{center}
\end{table}



\authorcontributions{All authors contributed to the writing of the paper. P.G., S.V., and M.D. conceptualized the paper. P.G. prepared the draft and visualizations. Review and editing were performed by S.V. and M.D. Funding acquisition and administration was done by S.V. and M.D.}

\funding{This research was funded by a W.M. Keck Foundation grant and is currently funded by NSF/ATI  grant 1711377, NASA/APRA grant 80NSS18K0242, and NASA Earth and Space Science Fellowship ASTRO18F-0085. }

\acknowledgments{The authors acknowledge  J. Bland-Hawthorn, S. Leon-Saval,  C. Betters, and B. Norris at Univ of Sydney for insights on various photonic instrumentation at Univ of Sydney. We also acknowledge  S. Ellis at AAO and J. Bryant,  A. Ariola, and  G. Douglas at Macquarie University for extensive discussion on current astrophotonic instrumentation at Australian National Fabrication Facility. The authors are very grateful to members of The Maryland Astrophotonics Laboratory  T. Zhu (former),  Y. Meng (former), Y. Hu, S. Xie, and J.~Zhan for lending their expertise in various photonic devices.}

\conflictsofinterest{The authors declare no conflict of interest.} 
\reftitle{References}


\end{document}